\documentclass{PoS}

\newcommand{\ba}{\begin{align}}
\newcommand{\ea}{\end{align}}
\newcommand{\beq}{\begin{eqnarray}}
\newcommand{\eeq}{\end{eqnarray}}

\title{Improved Maximum Entropy Method with an Extended Search Space}

\ShortTitle{Improved Maximum Entropy Method with an Extended Search Space}

\author{\speaker{A. Rothkopf}\\
        Albert Einstein Center for Fundamental Physics, Institute for Theoretical Physics, University of Bern, 3012 Bern, Switzerland\\
        E-mail: \email{rothkopf@itp.unibe.ch}}

\abstract{
We report on an improvement to the implementation of the Maximum Entropy Method (MEM). It amounts to departing from the search space obtained through a singular value decomposition (SVD) of the Kernel. Based on the shape of the SVD basis functions we argue that the MEM spectrum for given $N_\tau$ data-points $D(\tau)$ and prior information $m(\omega)$ does not in general lie in this $N_\tau$ dimensional singular subspace. Systematically extending the search basis will eventually recover the full search space and the correct extremum. We illustrate this idea through a mock data analysis inspired by actual lattice spectra, to show where our improvement becomes essential for the success of the MEM. To remedy the shortcomings of Bryan's SVD prescription we propose to use the real Fourier basis, which consists of trigonometric functions. Not only does our approach lead to more stable numerical behavior, as the SVD is not required for the determination of the basis functions, but also the resolution of the MEM 
becomes independent from the position of the reconstructed peaks.
}

\FullConference{The XXX International Symposium on Lattice Field Theory - Lattice 2012,\\
		June 24-29, 2012\\
		Cairns, Australia}

\begin{document}

\section{Introduction}

Lattice QCD is a central pillar for the non-perturbative investigation of the strong force. One important aspect is that its formulation in Euclidean time is amenable to Monte Carlo simulations at arbitrary $T$, i.e. also close to a phase transition. Computational advances over the last decade have furthermore made it possible to consistently probe the properties of e.g. the high temperature state of matter, the quark-gluon plasma, with staggered dynamical quarks in the continuum limit \cite{Borsanyi:2010bp} and to deploy the even more costly fermion formulations, such as Wilson or domain wall \cite{Cheng:2012iy}.

One particular area, which directly benefits from the availability of non-perturbative information about the QCD medium is the study of heavy quarkonia and its proposed melting \cite{Matsui:1986dk} in relativistic heavy ion collisions. In these studies the dynamics of a pair of a heavy quark and anti-quark is formulated in an effective field-theory framework (for a brief overview see e.g. \cite{Rothkopf:2012et} and references therein), which allows to make a close connection to the physics of the medium by relating the non-relativistic static potential governing the dynamics of the heavy $Q\bar{Q}$ to the real-time rectangular Wilson loop $W_\square(r,t)=\Big\langle {\rm exp}\Big[-i\int_\square dx^\mu A_\mu\Big] \Big\rangle$.

Unfortunately, formulating lattice QCD in imaginary time deprives us from direct access to dynamical information such as real-time observables. Field theory however provides a handy concept to relate imaginary and real-time quantities, through the use of spectral functions. These positive definite quantities $\rho(\omega)$ are connected to the observables $D(\tau)$ via an integral transform 
\begin{equation}
 D(\tau)=\int_{-\infty}^{\infty} K(\tau,\omega) \rho(\omega) d\omega\label{Eq:ConvolCont}
\end{equation}
with analytically known Kernel $K(\tau,\omega)$. Since all time dependence is explicitly located in the function $K(\tau,\omega)$ the analytic continuation $\tau\to it$ is straight forward, once the values of $\rho(\omega)$ are determined. (In the following we will focus on the exponentially damped Kernel $K(\tau,\omega)=e^{-\omega\tau}$).

In practice one obtains from simulations of lattice QCD a noisy estimate of the observable $D(\tau)$ at a discrete number of $N_\tau$ datapoints $D_i$. Extracting a function $\rho_l$ at $N_\omega\gg N_\tau$ frequencies from such a limited data-set via the inversion of
\begin{equation}
D_i = \sum_{l=1}^{N_\omega} K_{il}\; \rho_l\; \Delta\omega_l \label{Eq:ConvolDisc}
\end{equation}
is an ill-defined problem. How to give meaning to such a task is the central question elucidated by the Maximum Entropy Method \cite{Bryan:1990,Jarrell:1996,Nakahara:1999vy,Asakawa:2000tr}. This branch of Bayesian inference generalizes and ameliorates the concept of $\chi^2$ fitting by emphasizing the use of so called prior information. I.e. one introduces a function $m(\omega)$ which per definition is the correct spectral function in the absence of measured data\footnote{A complication exists in that since our prior information is usually derived from perturbation theory at high energies, there is essentially no information available on the low $\omega$ part of the spectrum, we are interested in. Thus we already know that the prior function used is not the correct solution in the absence of data.}.

From Bayes theorem we learn that the probability of a test spectral function to be the correct spectral function \vspace{-0.25cm}
\begin{equation}
P_{\rm MEM}[\rho|D,m]= \frac{ P[D|\rho]P[\rho|m] } { P[D|m] } \label{MEM:Bayes}
\end{equation}
for given $N_\tau$ measured data $D(\tau_i)=D_i$ and $N_\omega$ prior information $m(\omega_l)=m_l$ can essentially be written as a product of the likelihood probability\vspace{-0.25cm}
\begin{equation}
P[D|\rho]=e^{-{\cal L}}={\rm exp}\Big[ -\frac{1}{2} \sum_{i,j=1}^{N_\tau} (D_i-D^\rho_i)C_{ij}^{-1} (D_j-D^\rho_j) \Big],\label{Eq:LikelihodProb}
\end{equation}
with $C_{ij}$ denoting the covariance matrix and the prior probability\vspace{-0.25cm}
\begin{equation}
P[\rho|m]=e^{\alpha {\cal S}} = {\rm exp}\Big[ \alpha \sum_{l=1}^{N_\omega} \Big( \rho_l-m_l-\rho_l{\rm log}[\frac{\rho_l}{m_l}]\Big)\Big]. \label{Eq:PriorProb}
\end{equation}
The parameter $\alpha$ introduced in eq.(\ref{Eq:PriorProb}) is selfconsistently determined and does not enter the final result \cite{Jarrell:1996,Nakahara:1999vy}. The task to perform is to find the global extremum of eq.(\ref{MEM:Bayes}),
\begin{equation}
\left. \frac{\delta}{\delta \rho} P_{\rm MEM}[\rho|D,m] \right|_{\rho=\rho_{\rm MEM}}=0\label{MEM:Optimize}
\end{equation}
which is in general very costly, since it involves an $N_\omega$ dimensional search space. Note that it has been proven in \cite{Asakawa:2000tr} that the solution of eq.(\ref{MEM:Optimize}) is unique if it exists. This statement does neither depend on a particular parametrization of the test spectral function $\rho(\omega)$ nor on the number of available data-points. Indeed in the extreme case of zero data-points the solution spectrum of the MEM would be the prior function.

In maximizing $P_{\rm MEM}[\rho|D,m]$ the likelihood and the prior term compete. Pure $\chi^2$ fitting is under-determined, so an infinite number of solutions exist, which all reproduce the measured data within the errorbars. The prior then selects from within these solutions the one that is closest to $m(\omega)$ while still respecting the constraints from the data. In essence one regularizes the $\chi^2$ fitting, with parts of the spectrum being determined by the data, parts of it being fixed by the prior. Thus to find out which part of the spectrum is actually encoded in the lattice QCD correlator, we need to redo the MEM with several distinct prior function to observe which part of $\rho^{MEM}(\omega)$ is invariant under a change of $m(\omega)$.

To carry out the maximization procedure in eq.(\ref{MEM:Optimize}), the state of the art implementation of the MEM relies on the singular value decomposition prescription introduced by Bryan \cite{Bryan:1990}. (For an alternative implementation without the use of the SVD see \cite{Jakovac:2006sf}) In the following we will argue that the global extremum of eq.(\ref{MEM:Bayes}) does not in general lie in his $N_\tau$ dimensional singular search space but one needs to systematically approach the full $N_\omega$ dimensional space.
\vspace{-0.3cm}
\section{Extended Search Space}
Let us quickly revisit Bryan's argument by inserting eq.(\ref{Eq:LikelihodProb}) and eq.(\ref{Eq:PriorProb}) into eq.(\ref{MEM:Optimize}) and making explicit the positive definiteness of the spectrum by using $\rho_l=m_l{\rm exp}[a_l]$, which together with the SVD of $K^t=U\Sigma V^t$ yields\vspace{-0.3cm}
\begin{equation}
 -\alpha \vec{a} =  K^t \vec{\frac{d{\cal L}}{dD^\rho}}=  U \Sigma V^t  \vec{\frac{d{\cal L}}{dD^\rho}}. \label{Eq:DefSVDsp}
\end{equation}
Since $\Sigma$ is a $N_\omega\times N_\omega$ diagonal matrix with only $N_\tau$ entries different from zero, Bryan concluded that eq.(\ref{Eq:DefSVDsp}) restricts the values of $\vec{a}$ to the space spanned by the first $N_\tau$ columns of $U$. We check the validity of this claim by numerically investigating the behavior of the corresponding basis functions $U^i(\omega_l)=U_{il}$ ($i=1,\ldots,N_\tau$) of Bryan's search space. 

\begin{figure}[t!]
\centering
 \includegraphics[scale=0.33,angle=-90,clip,trim = 0mm 0mm 0mm 45mm]{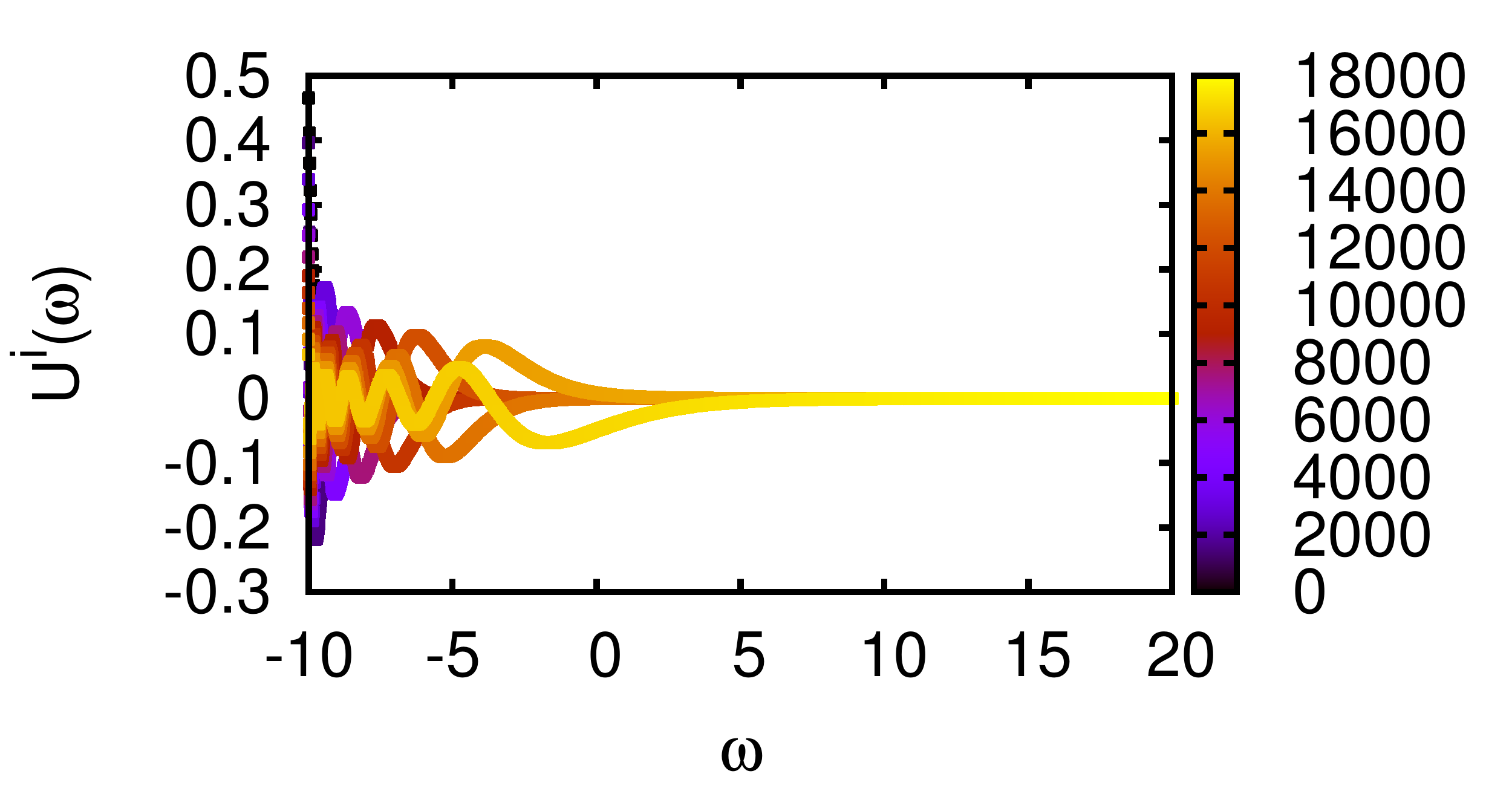}\hspace{0.5cm}
 \includegraphics[scale=0.33,angle=-90,clip,trim = 0mm 0mm 0mm 45mm]{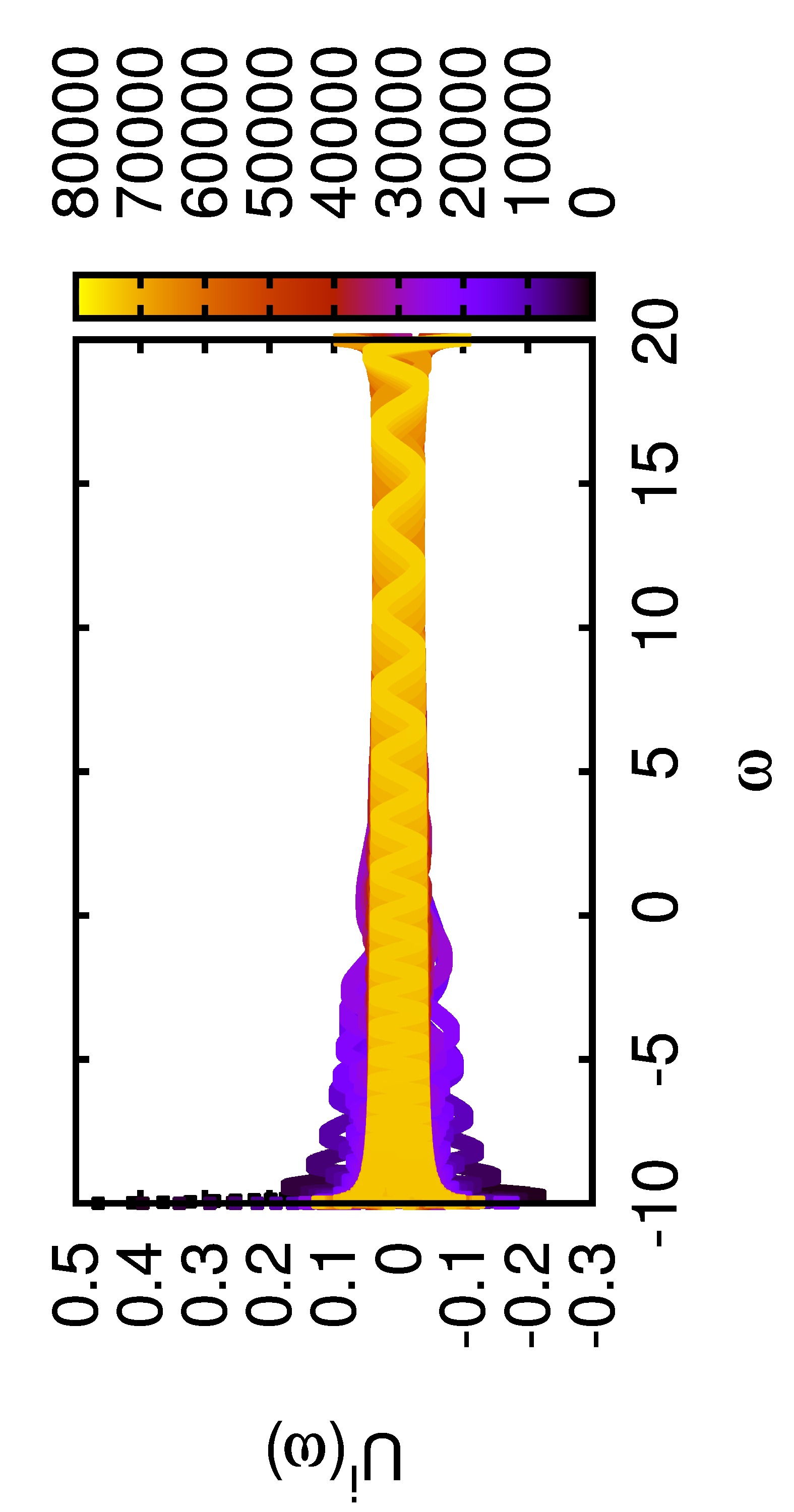}
 \caption{(left) The first twelve basis functions from the exponential Kernel $e^{-\omega\tau}$ using $N_\omega=1500$ to discretize the interval $\omega\in[-10,20]$, while $\tau\in(0,6.1]$ is discretized in $N_\tau=12$ steps. Note that at positive frequencies above $\omega=5$ no structure remains, so that reconstruction of peaks becomes very difficult. (right) The first 50 columns of the matrix U as comparison, where the oscillations are visible over the whole frequency range.}\label{Fig:SVDBasisFuncs}
\end{figure}

In Fig.\ref{Fig:SVDBasisFuncs} we show the basis functions of the Kernel $K(\omega,\tau)=e^{-\omega\tau}$, where we have chosen to discretize eq.(\ref{Eq:ConvolCont}) over a finite interval $\omega\in[\omega_{\rm min}=-10,\omega_{\rm max}=20]$. The functions $U^i(\omega_l)$ exhibit a common feature, i.e. they posses an oscillating part, which starts from $\omega_{\rm min}$ and extends up to a certain $\omega_{\rm osz}$. Beyond this frequency the functions rapidly damp to zero. Hence in the example shown, with the choice $\omega_{\rm min}=-10$, the first $N_\tau=12$ basis function do not allow the reconstruction of peak structures in the region $\omega>\omega_{\rm osz}\simeq0$. The conceptual problem of Bryan's SVD basis lies in the fact that the choice of $\omega_{\rm min}$ is completely arbitrary. As long as the resolution of the frequency interval stays the same $\Delta\omega=\frac{\omega_{\rm max}-\omega_{\rm min}}{N_\omega}$, extending the frequency range to smaller and smaller $\omega_{\rm min}$ is admissible. 
Decreasing $\omega_{\rm min}$ does not change the length of the oscillatory part it just shifts the basis functions to smaller and smaller frequencies. At some point, the MEM in the singular subspace cannot reproduce the measured datasets, since it does not have any peak structures left at positive frequencies, while in the full search space the data can be reconstructed without problem.

In other words, no matter how many datapoints have been measured it is always possible to find a $\omega_{ \rm min}$ so that e.g. the positive frequency range cannot be resolved. We conclude that the success of Bryan's MEM prescription depends on the arbitrary choice of $\omega_{ \rm min}$. The proof for existence and uniqueness of the MEM solution on the other hand does not require a certain parametrization of the spectral function and thus is valid for the full $N_\omega$ dimensional space. This tells us that the SVD space artificially restricts the optimization without guaranteeing the presence of the correct global extremum.

Thus as a first step we propose to systematically enlarge Bryan's search space \cite{Rothkopf:2011ef} to approach the full $\mathbb{R}^{N_\omega}$, which has to contain the correct solution. To this end we parametrize the spectral function with $N_{\rm ext}>N_\tau$ parameters $b_j$ \vspace{-0.2cm}
\begin{equation}
\rho_l=m_l {\rm exp}\Big[\sum_{j=1}^{N_{\rm ext}} \,U_{lj}\, b_j\Big].
\end{equation}
As a rule of thumb we can stop increasing $N_{\rm ext}$ once the value of $P_{\rm MEM}[\rho|D,m]$ stops to grow after adding another basis vector.

\begin{figure}[t!]
 \includegraphics[scale=0.28,angle=-90]{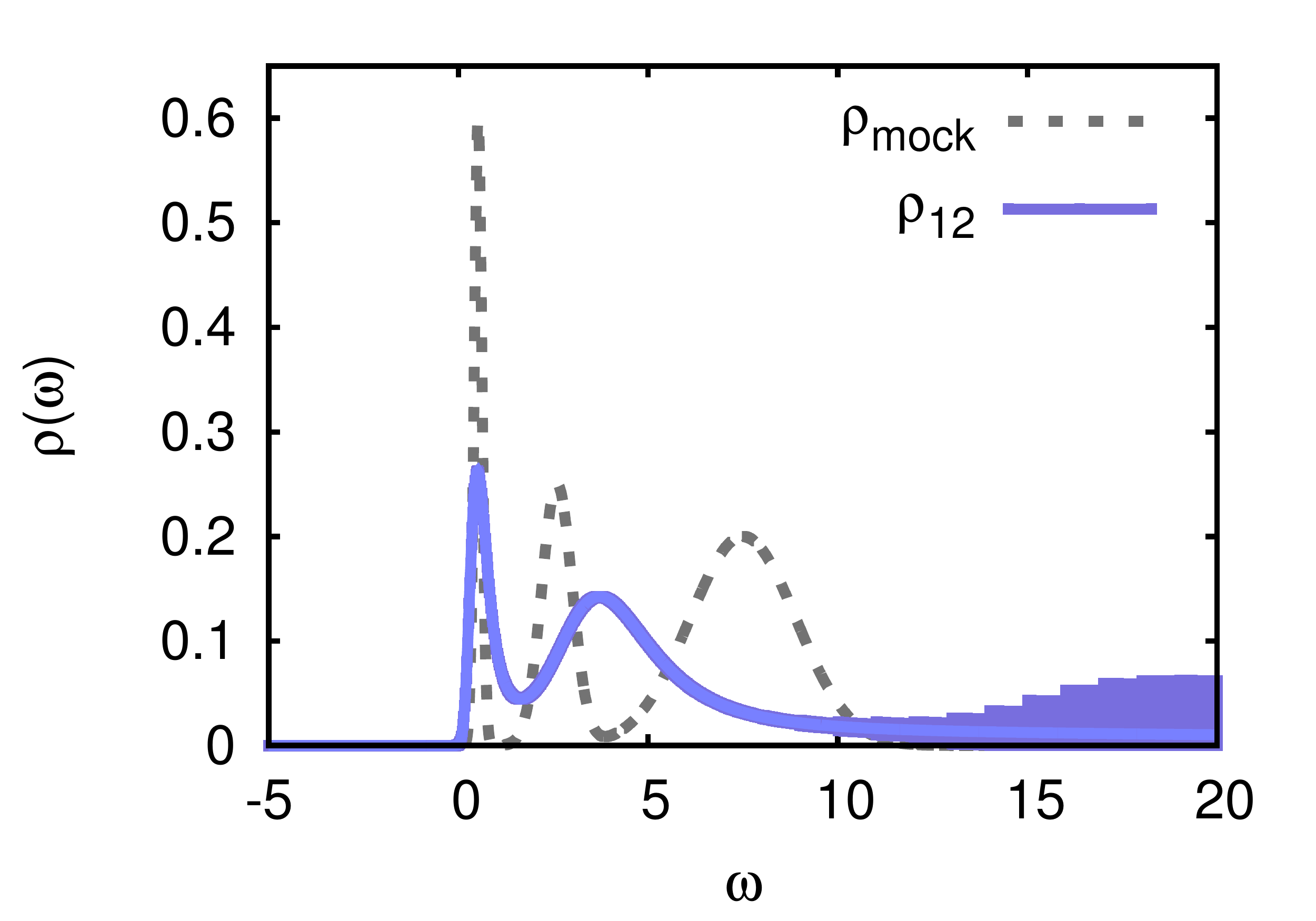}
 \includegraphics[scale=0.28,angle=-90]{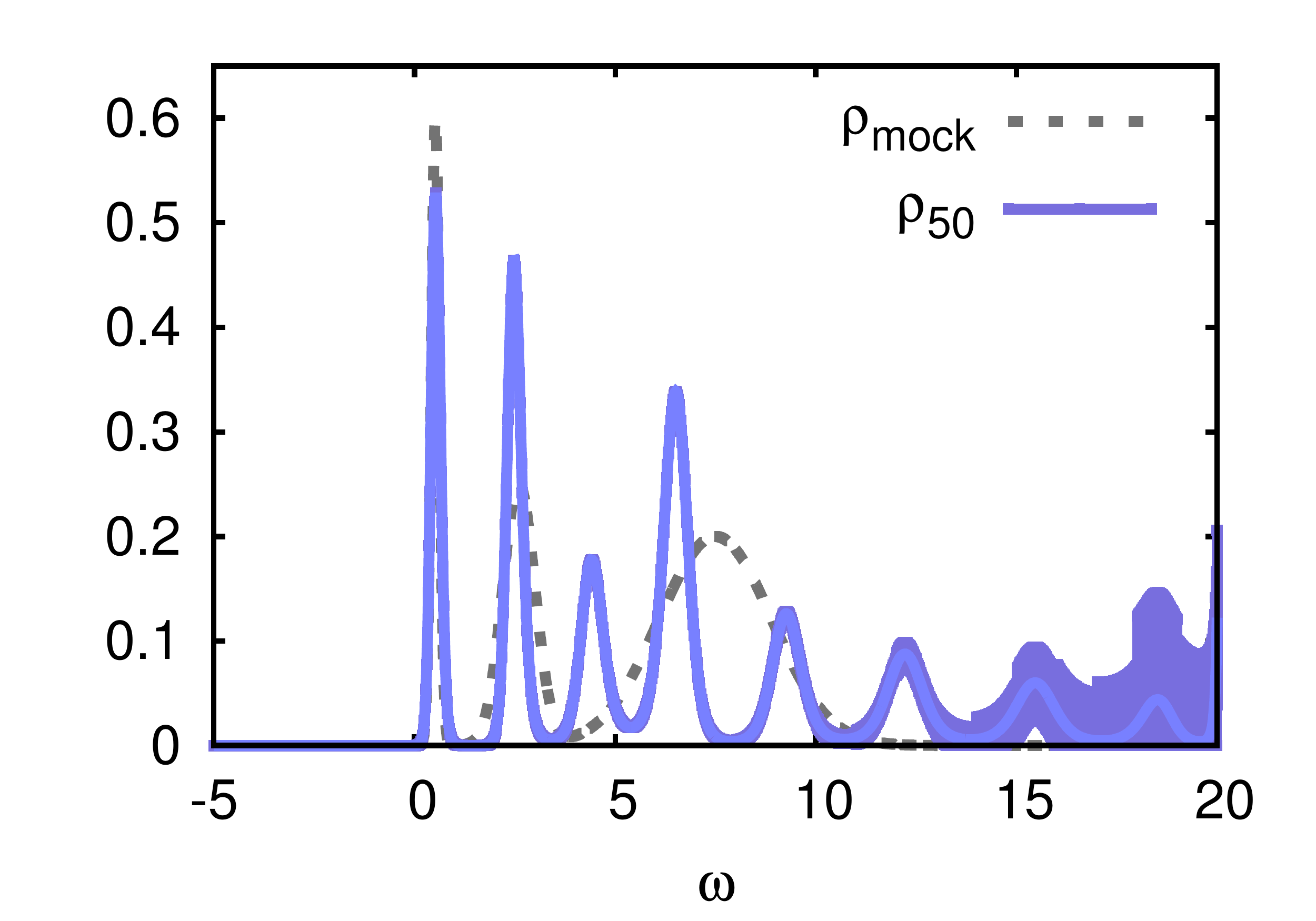}
 \caption{(left) Reconstruction of the mock spectral function (gray dashed) using Bryan's prescription on $N_\tau=12$ datapoints. Note that even the lowest lying peak is washed out and higher lying structures are not captured at all. (right) Based on the same data, the MEM in the extended search space $N_{\rm ext}=50$ succeeds in reconstructing the lowest lying peak. Wiggling behavior at larger $\omega$ is an artifact of the chosen prior and can be identified as such by variation of $m(\omega)$}\label{Fig:ReconstrSpec}
\end{figure}

To illustrate how the extension of the search space can improve the MEM reconstruction, we present a mock data analysis, inspired by the spectra encountered in \cite{Rothkopf:2011db}. A discrete mock spectrum is calculated from an analytic expression containing the sum of four Gaussian peaks as shown through the gray dashed lines in Fig.\ref{Fig:ReconstrSpec} (One exponentially small peak is present at negative frequencies $\omega=-2.6$). From it, a set of $N_\tau=12$ data-points is generated via eq.(\ref{Eq:ConvolDisc}), which after being perturbed by Gaussian noise with $\sigma_j=0.01 j D^{\rm ideal}_j, \quad j\in[1,\cdots,N_\tau]$ is fed to the MEM code as input. For the prior we choose the function $m(\omega)=\frac{1}{\omega-\omega_m}$, so that $m(\omega_{ \rm min})=0.01$.

On the left of Fig.\ref{Fig:ReconstrSpec} we show that according to Bryan's SVD basis the spectrum is not well reproduced. The position and width of even the lowest lying peak is not captured adequately, structure at higher $\omega$ is completely absent. Unsurprisingly the probability of this solution is very small ($Q=log[P_{\rm MEM}[\rho|D,m]]\simeq-10^4$). In contrast, after extending the basis to $N_{\rm ext}=50$, we are able to reconstruct from the same $N_\tau=12$ dataset the lowest lying peak to a much better degree ($Q=-5.5$) with both position and width being closer to the mock spectrum. This result is a direct counterexample to the claim that the global extremum of $P_{\rm MEM}[\rho|D,m]$ will always lie in the singular subspace.

At higher frequencies a lot of wiggly structures appear on the right of Fig.\ref{Fig:ReconstrSpec}, which are an artifact of the choice of prior function. These can be identified as such by repeating the MEM reconstruction with different prior functions $m(\omega)$ and observing that only the lowest lying peak remains invariant. 
\vspace{-0.2cm}

\section{Fourier Basis}

\vspace{-0.2cm}
We have seen that the MEM solution, i.e. the global maximum of the functional $P_{\rm MEM}[\rho|D,m]$ does not in general lie in the singular subspace obtained from the first $N_\tau$ columns of the SVD matrix $U$. We use this freedom to propose a completely different basis for the search space, which cures the central disadvantage of Bryan's SVD basis functions. Our choice decouples the success of the reconstruction from the the position of the spectral features relative to $\omega_{ \rm min}$

\begin{figure}[t]
 \includegraphics[scale=0.3,angle=-90]{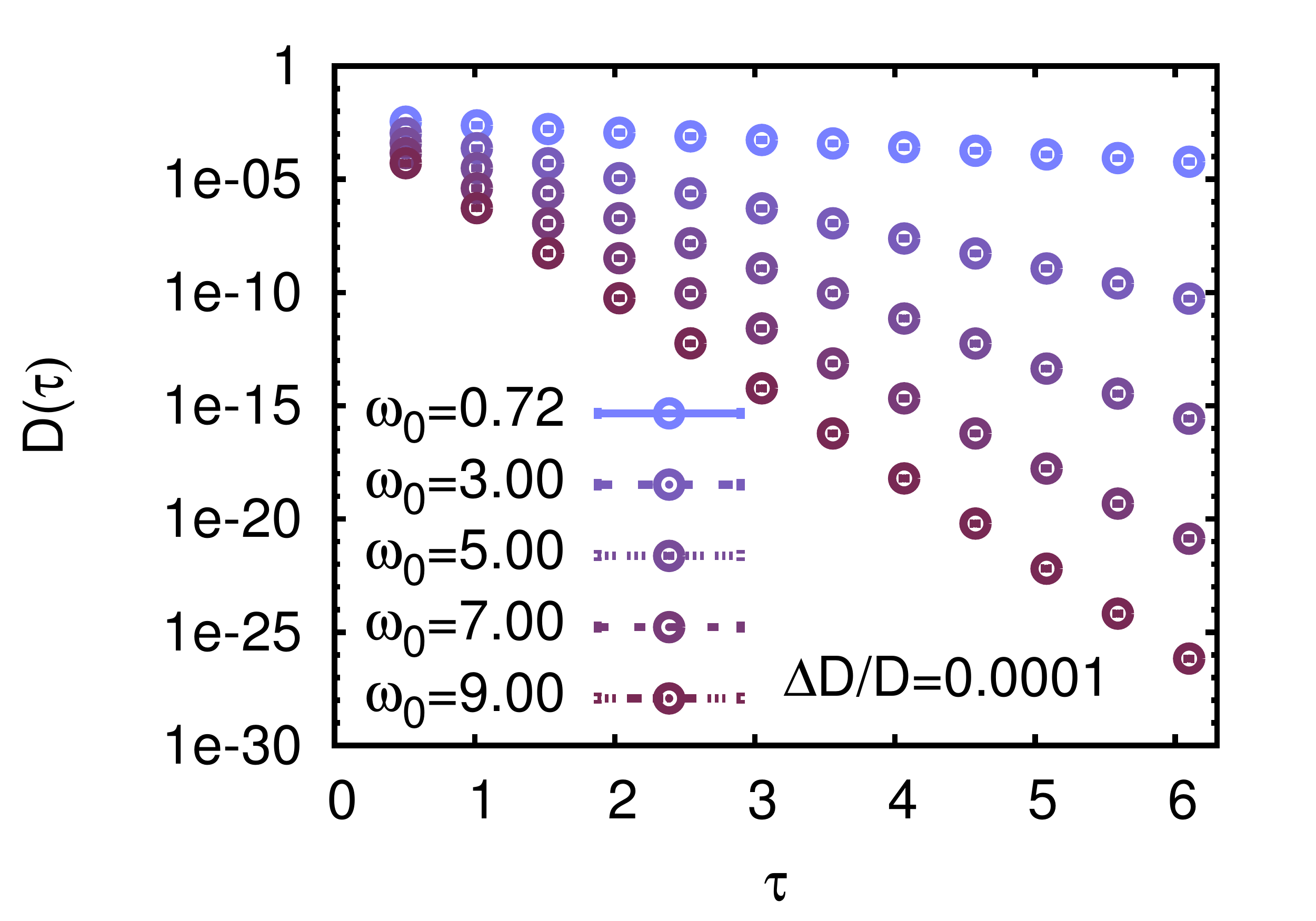}
 \includegraphics[scale=0.3,angle=-90]{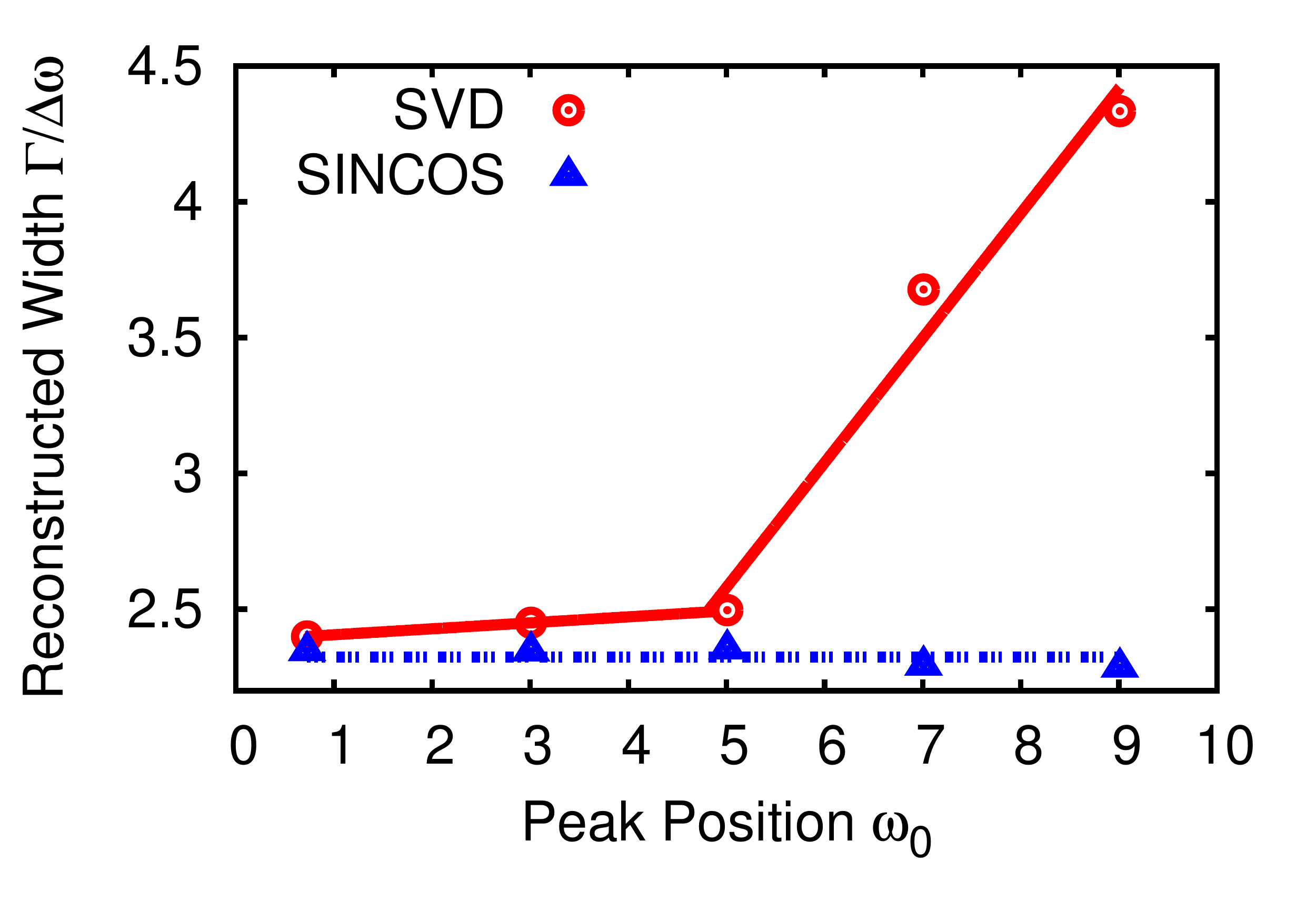}
 \caption{(left) $N_\tau=12$ mock data coming from a single delta peak positioned at increasing frequency $\omega_0$. We choose to perturb the data with Gaussian noise, such that a constant relative error ensues. (right) Reconstructed peak width versus peak position. The SVD basis yields acceptable results while the peak lies within the region, where oscillations exist in the basis functions. Once in the damping region $\omega>5$ the resolution of the reconstructed width deteriorates. (Note that here $\omega\in[-5,20]$ so that the SVD basis functions oscillate until $\omega_{\rm osc}\simeq5$, different from the ones in Fig.1) In contrast, the Fourier basis provides a position independent resolution, which even slightly improves due to the better absolute errors in the data at larger $\omega_0$.}\label{Fig:FourierBasis}
\end{figure}

Based on the real Fourier basis,we can parametrize the spectral function in closed form
$$
\rho_l=m_l {\rm exp} \Big[ b_1 + \sum_{j=1}^{N_{\rm ext}/2}b_{2j} {\rm sin}[(\omega_l-\omega_{\rm min}) j] + \sum_{j=1}^{N_{\rm ext}/2}b_{2j+1} {\rm cos}[(\omega_l-\omega_{\rm min}) j]\Big]
$$
First note that since we have an analytic formula available in this case, there is no need to perform the SVD to determine the basis functions. Especially for large numbers of $N_{\rm ext}>50$ this translates into a numerically more robust determination of the spectrum, since for the SVD to accurately reproduce the highly oscillating parts of the basis functions a very large numerical precision is usually required.  

The main feature of this parametrization is that the choice of $\omega_{ \rm min}$ becomes irrelevant to the success of the reconstruction. To illustrate this point we perform several mock data analyses based on single delta peaks placed at different frequencies $\omega_0$. The data obtained from these spectra (see Fig.\ref{Fig:FourierBasis} left) shows a simple exponential falloff and is perturbed by Gaussian noise with a $\sigma_i=0.0001D^{ideal}_i$ before being fed to the MEM as input.

As expected from the finite number of datapoints and the finite error present, the single delta peak is not fully resolved (i.e. $\Gamma/\Delta\omega>1$), no matter where it is positioned at positive frequencies. Fig. \ref{Fig:FourierBasis} shows the reconstructed width versus the peak position, which allows us to observe that the MEM resolution deteriorates in the SVD basis for peaks at the higher frequency positions. If the peaks lie within the oscillating region of the SVD basis functions their reconstruction only suffers mildly from the peak position, but once we position the peak in the purely damped regime, the artificially induced width in the reconstruction increases strongly.

The Fourier basis behaves much more favorably, as it does not show any decrease in resolution with change in the peak position. Note that the width of the reconstruction even becomes slightly smaller for peaks at high frequencies, since there the absolute error in the mock data is very small. Hence we have found a remedy to the central problem connected with Bryan's basis, i.e. the Fourier basis allows us to reconstruct peak structures independently from the choice of $\omega_{ \rm min}$.

\section{Summary and concluding remarks}

We have shown by an inspection of the basis functions used in Bryan's approach that in general the MEM solution does not lie in the $N_\tau$ dimensional search space obtained by an SVD of the transpose kernel. Since the uniqueness of a possible solution in the full $N_\omega$ dimensional space was proven on general grounds in \cite{Asakawa:2000tr}, we proposed as a first step to systematically enlarge Bryan's search space by including more and more columns of the SVD matrix $U$. The reconstructed spectra, presented in Fig.\ref{Fig:ReconstrSpec}, which were based on mock data inspired by QCD correlators found in \cite{Rothkopf:2011db}, represent an example of where our improvement becomes crucial for the success of the MEM.

In order to improve the MEM further, we introduced the Fourier basis to parametrize the spectral function based on trigonometric functions. The availability of an analytic expression in this case renders the SVD superfluous. Hence the numerical results are more stable, as the SVD requires a large numerical precision to correctly determine the highly oscillating features of its basis functions. Deploying the Fourier basis to reconstruct single delta peak mock-spectra, we find that the resolution of the result does not depend on the peak position, in contrast to the SVD basis, which shows a deteriorating resolution for peaks at increasing frequencies. 

The source code of the MEM program used to carry out all of the mock data analyses presented here can be found on-line at www.scicode.org/ExtMEM together with a short manual and instructions for compilation.

For support and inspiration the author is indebted to T.~Hatsuda and S.~Sasaki. A.R. would also like to thank O.~Kaczmarek, S.Y.~Kim, J.-I.~Skullerud  and P.~Petreczky for many fruitful discussions. The speaker is grateful for generous access to computational resources through the DFG-Heisenberg group of Y.~Schr\"oder at Bielefeld University and acknowledges partial support by the BMBF under project {\em Heavy Quarks as a Bridge between Heavy Ion Collisions and QCD}. In addition, the attendance at this conference was partially supported by the ``Nachwuchsf\"orderung'' program of the faculty of natural sciences at the University of Bern.

\end{document}